\documentclass[12pt]{article}
\usepackage{epsfig}
\usepackage{graphicx}

\def\beq{\begin{equation}}
\def\eeq#1{\label{#1}\end{equation}}
\def\eeqn{\end{equation}}

\def\beqa{\begin{eqnarray}}
\def\eeqa#1{\label{#1}\end{eqnarray}}
\def\eeqan{\end{eqnarray}}

\let\bar=\overbar

\def\Dslash{\not{\hbox{\kern-4pt $D$}}}
\def\dslash{\not{\hbox{\kern-2pt $\del$}}}

\def\msb{{\bar{\ssstyle M \kern -1pt S}}}

\usepackage{fancyhdr,graphicx}
\def\Title#1{\begin{center} {\Large {\bf #1} } \end{center}}

\begin{document}

\Title{Two types of glitches in a solid quark star model$^{\dag}$}

\footnotetext{$^{\dag}$ This brief report is based on the article, Zhou E. P., Lu J. G., Tong H. $\&$ Xu R. X., 2014, MNRAS, 443, 2705.

\hspace*{5mm}$^{\bigtriangleup}$ zhouenpingz715@sina.com\\

}
\bigskip\bigskip


\begin{raggedright}

{\it 
Enping Zhou$^{1}$

\bigskip
$^{1}$Department of Astronomy, Peking University, Beijing, 100871, China
\bigskip

}

\end{raggedright}

\section{Introduction}

A pulsar is a rapidly spinning magnetized compact star.
The highly collimated beam of radiation of the star is rotated into and out of the 
line of sight of a distant observer as the star itself rotates around a 
fixed axis, manifesting a periodic sequence of pulsations.
Hence, pulsars are thought to be the accurate clocks in the Universe.

During pulsar observations, a sudden increase in pulsar's spin frequency $\nu$ (i.e. glitch) is discovered. The
observed fractions $\Delta\nu/\nu$ range between $10^{-10}$ and $10^{-5}$ (Yu et al. 2013).
In neutron star models, pulsars are thought to be a fluid star with a thin
solid shell. The physical mechanism of glitch is believed to be the coupling
and decoupling between outer crust (rotates slower) and the inner superfluid
(rotates faster) (Anderson $\&$ Itoh 1975; Alpar et al. 1988). However, the absence of evident energy
release during even the largest glitches ($\Delta \nu/\nu \sim 10^{-6}$)
of Vela pulsar is a great challenge to this glitch scenario (G$\mathrm{\ddot{u}}$rkan et al. 2000; Helfand et al. 2001).
Recently, the glitches detected from AXP/SGRs (anomalous X-ray
pulsars/soft gamma repeaters), which are usually accompanied with energy release
 (Kaspi et al. 2003; Tong $\&$ Xu 2011; Dib $\&$ Kaspi 2014) become another challenge to the previous glitch models.
\section{Two types of glitches}
In our model, glitches can be divided into two types depending on the energy released.
According to our calculations, these two types of glitches might be induced by two
types of starquakes of solid stars, i.e. the bulk variable starquake and bulk invariable starquake.

The bulk variable starquake is induced by the accretion. 
Due to the accretion, the mass of the star increases, and the radius will change as well.
However, as a solid star, the pulsar will gain elastic energy to resist this change in shape.
A sudden collapse will happen when this energy is too large for the solid structure to stand.
With a typical pulsar and glitch parameter, the estimated energy released in this way coincides with the observed value during the glitches of AXP/SGRs.

The other type of starquake occurs without a bulk variation, and it's induced by the spinning down of the pulsar.
The pulsars will spin down due to the magnetic dipole radiation. Stars made of perfect fluid will change its ellipticity
and deform like a Maclaurin ellipsoid when it spins down. However, for a solid star, the shearing force will resist this change
in ellipticity and gain additional elastic energy. A starquake will happen when the shearing force exceeds a critical value.
The moment of inertia will decrease and the pulsar will spin up. With some available parameters, we can reproduce the glitch properties,
 i.e. the glitch amplitude and the intervals between glitches, with very little energy released.

Fig. 1 show our calculating result of the two types of starquakes.
For same glitch amplitude, the energy released during a bulk variable
will be about $10^7$ times larger than that of bulk invariable starquake.
Besides the energy release estimation, there are also some other hints showing that the glitch 
on AXP/SGRs and the glitch on Vela might be bulk variable starquake and bulk invariable starquake, respectively.
Some observation hints the possibility of accretion on AXP/SGRs, which is the key to induce a 
bulk varible starquake. And on the other hand, as a younger pulsar,
the spin of Vela is much faster than AXP/SGRs, so that the role of stellar ellipticity is much more 
important for the evolution of Vela-like pulsars.

In neutron star models, a pulsar is thought to be a fluid star with a thin solid shell.
In fact, so far only quark star and quark cluster star model develop a solid star model.
Then the two types of glitches may be an implication that the pulsar
is composed by quark matter or quark cluster matter.

\begin{figure}[htb]
\includegraphics[width=0.6\textwidth]{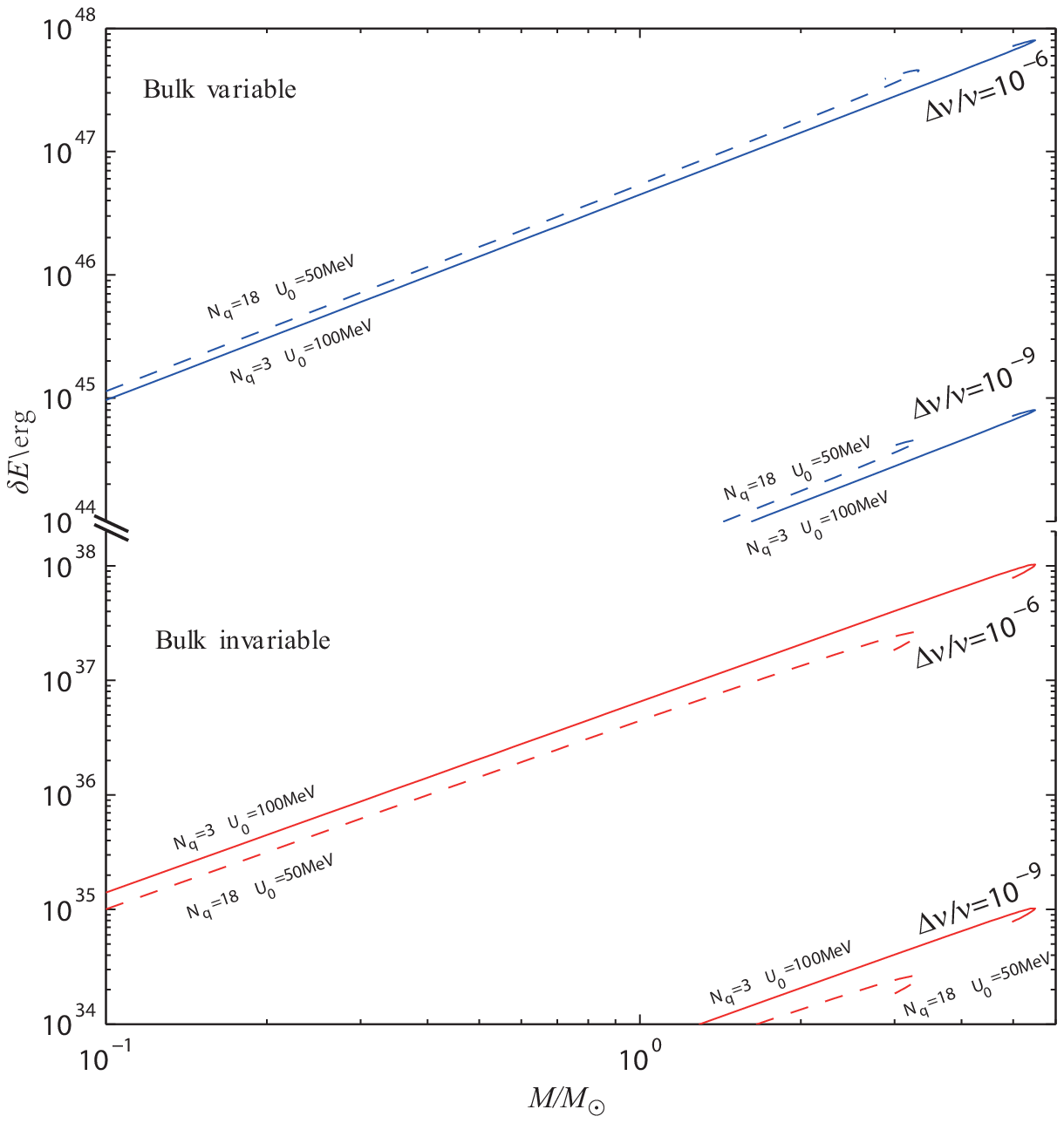}
\caption{The total energy release during the bulk-variable glitches and bulk-invariable
glitches with amplitudes of $10^{-6}$ and $10^{-9}$.
The Lennard-Jones interaction is applied as an approximation to work out
the mass-radius relation (Lai $\&$ Xu 2009).
There are two main factors in this approximation: the number of quarks in one
cluster ($N_{\mathrm{q}}$) and the depth of the potential($U_{0}$).
The case of 3-quark clusters with potential of 100MeV (solid lines) and 18-quark
clusters with potential of 50MeV (dashed lines) are considered.
It's also worth noting that the energy release during a bulk-invariable
glitch is related to the time intervals between two glitches.
In this calculation the glitch is thought to happen once per month and the
spin down power is calculated according to the observational data of Vela.}
\label{fig:Mott}
\end{figure}

\subsection*{Acknowledgement}

We express our thanks to the organizers of the CSQCD IV conference for providing an 
excellent atmosphere which was the basis for inspiring discussions with all participants.
We have greatly benefitted from this.

\end{document}